\documentclass[fleqn,usenatbib]{mnras}
\usepackage{newtxtext,newtxmath}
\usepackage[T1]{fontenc}
\usepackage{ae,aecompl}
\usepackage{graphicx}	% Including figure files
\usepackage{amsmath}	% Advanced maths commands
\usepackage{amssymb}	% Extra maths symbols
\usepackage{subfigure}
\usepackage{multicol}  
\usepackage{multirow} 
\usepackage{verbatim}
\usepackage{epsfig}
\usepackage{lscape}
\usepackage{natbib}

\newcommand{\GG}[1]{}
\newcommand{\kms}{{\rm \,km\,s^{-1}}}
\newcommand{\Mpc}{\,{\rm Mpc}}
\newcommand{\kpc}{\,{\rm kpc}}

\newcommand{\Mstar}{M_{*}}
\newcommand{\Mhalo}{M_{\rm halo}}
\def\vc#1{{\mbox{\boldmath$#1$\unboldmath}}}

\title[Stellar orbit fractions of the TNG100 galaxies] {A study of stellar orbit fractions: simulated IllustrisTNG galaxies compared to CALIFA observations} 
\author[Xu et al.] {Dandan Xu$^{1}$\thanks{E-mail: dandanxu@tsinghua.edu.cn}, Ling Zhu$^{2}$\thanks{E-mail: lzhu@shao.ac.cn}, Robert Grand$^{3}$, Volker Springel$^{3}$, Shude Mao$^{4,5}$, \and Glenn van de Ven$^{6}$, Shengdong Lu$^{5}$, Yougang Wang$^5$, Annalisa Pillepich$^{7}$, \and Shy Genel$^{8,9}$, Dylan Nelson$^{3}$, Vicente Rodriguez-Gomez$^{10}$, R{\"u}diger Pakmor$^{3}$, \and Rainer Weinberger$^{11}$, Federico Marinacci$^{12}$, Mark Vogelsberger$^{13}$, \and Paul Torrey$^{14}$, Naiman Jill$^{11,15}$, Lars Hernquist$^{11}$
\\\\ $^{1}$ Institute for Advanced Studies and Tsinghua Center for Astrophysics, Tsinghua University, 100084, Beijing, China
\\ $^{2}$ Shanghai Astronomical Observatory, Chinese Academy of Sciences, 80 Nandan Road, Shanghai 200030, China
\\ $^{3}$ Max-Planck-Institut f{\"u}r Astrophysik, Karl-Schwarzschild-Str. 1, D-85748, Garching, Germany
\\ $^{4}$ Department of Astronomy and Tsinghua Center for Astrophysics, Tsinghua University, 100084, Beijing, China 
\\ $^{5}$ National Astronomical Observatories, Chinese Academy of Sciences, 20A Datun Road, Chaoyang District, Beijing 100101, China
\\ $^{6}$ Department of Astrophysics, University of Vienna, T{\"u}rkenschanzstrasse 17, 1180 Vienna, Austria
\\ $^{7}$ Max-Planck-Institut f{\"u}r Astronomie, K{\"o}nigstuhl 17, 69117 Heidelberg, Germany
\\ $^{8}$ Center for Computational Astrophysics, Flatiron Institute, 162 Fifth Avenue, New York, NY 10010, USA
\\ $^{9}$ Columbia Astrophysics Laboratory, Columbia University, 550 West 120th Street, New York, NY 10027, USA
\\ $^{10}$ Instituto de Radioastronom{\'i}a y Astrof{\'i}sica, Universidad Nacional Aut{\'o}noma de M{\'e}xico, A.P. 72-3, 58089 Morelia, Mexico
\\ $^{11}$ Harvard–Smithsonian Center for Astrophysics, 60 Garden Street, Cambridge, MA 02138 
\\ $^{12}$ Department of Physics \& Astronomy, University of Bologna, via Gobetti 93/2, 40129 Bologna, Italy      
\\ $^{13}$ Department of Physics, Kavli Institute for Astrophysics and Space Research, Massachusetts Institute of Technology, MA 02139, USA   
\\ $^{14}$ University of Florida, Department of Physics, 2001 Museum Rd., Gainesville, FL 32611, USA 
\\ $^{15}$ The School of Information Sciences at the
University of Illinois Urbana-Champaign, 501 E Daniel St, Champaign, IL 61820, USA } 
\date{\vspace{-5em}}

\begin{document}
\pagerange{\pageref{firstpage}--\pageref{lastpage}} \pubyear{2019}
\maketitle
\label{firstpage}
\begin{abstract}

Motivated by the recently discovered kinematic ``Hubble sequence'' shown by the stellar orbit-circularity distribution of 260 CALIFA galaxies, we make use of a comparable galaxy sample at $z=0$ with a stellar mass range of $\Mstar/M_{\odot}\in[10^{9.7},\,10^{11.4}]$ selected from the IllustrisTNG simulation and study their stellar orbit compositions in relation to a number of other fundamental galaxy properties. We find that the TNG100 simulation broadly reproduces the observed fractions of different orbital components and their stellar mass dependencies. In particular, the mean mass dependencies of the luminosity fractions for the kinematically warm and hot orbits are well reproduced within model uncertainties of the observed galaxies. The simulation also largely reproduces the observed peak and trough features at $\Mstar \approx 1-2\times 10^{10}M_{\odot}$ in the mean distributions of the cold- and hot-orbit fractions, respectively, indicating fewer cooler orbits and more hotter orbits in both more- and less-massive galaxies beyond such a mass range. Several marginal disagreements are seen between the simulation and observations: the average cold-orbit (counter-rotating) fractions of the simulated galaxies below (above) $\Mstar \approx 6\times 10^{10}M_{\odot}$ are systematically higher than the observational data by $\la 10\%$ (absolute orbital fraction); the simulation also seems to produce more scatter for the cold-orbit fraction and less so for the non-cold orbits at any given galaxy mass. Possible causes that stem from the adopted heating mechanisms are discussed.

\end{abstract}

\begin{keywords}
  galaxies: haloes - galaxies: structure - dark matter.
\end{keywords}

\section{Introduction}
The orbit composition of a stellar system is a key concept in galactic dynamics (\citealt{1987gady.book.....B}). It develops under various dynamical processes and instabilities due to gravity. Put into a larger context, a galaxy's stellar orbit composition is an emergent phenomenon during a galaxy's assembly (\citealt{1978MNRAS.183..341W, 1983ApJ...266..516D, 2006MNRAS.368..623M, 2010ApJ...721.1878S}). It forms through many generations of star formation, the history of which is determined by a variety of conditions, from the channel and efficiency of cold gas replenishment which fuels star formation, to the location and strength of energy and matter feedback which regulate the next-generation of star formation.

These seemingly internal processes are also closely connected to a galaxy's large-scale environment, where neighbouring galaxies with different orbits, masses and gas contents may merge. During this process, existing stars may become kinematically hotter due to dynamical scattering, and stars accreted on radial orbits populate kinematically \textit{hot} orbits. Meanwhile, a new generation of star formation can be triggered when fresh cold gas is brought into the system (via wet mergers, \citealt{2006ApJ...645..986R}), which may result in kinematically \textit{cold} stellar orbits if gas accretion occurs preferentially on \textit{tangential} orbits. Mergers may also quench star forming due to intense starbursts and AGN activity triggered by nuclear gas inflows (\citealt{1989Natur.340..687H, 2005ApJ...620L..79S}).

These processes determine not only the star formation history and stellar orbit composition of a galaxy, but also other fundamental properties, such as its mass, size, colour and morphology. All these different aspects are inter-connected: the fraction of stars that possess kinematically cold (hot) orbits in a galaxy is found to correlate well with the fraction of stars that compose a disk (bulge) morphology of the galaxy (\citealt{2018MNRAS.479..945Z}); at a given mass, more spherical-shaped galaxies are generally found to be redder (due to dominance of older stellar populations) and smaller than their flatter disk-shaped counterparts (e.g., \citealt{2016arXiv161205560C, 2019MNRAS.483.4140R}).

But how do kinematically different orbital components emerge in the first place and how do they evolve in the context of a galaxy's secular evolution and hierarchical merging? What drives a galaxy to quench its star forming activity and transition from a spiral/disk galaxy, in which cold stellar orbits are prevalent, to an elliptical where hot stellar orbits dominate? During galaxy type transition, how do galaxy properties change and provide feedback to this process?

Thanks to the advent of cosmological hydrodynamic full-physics simulations of galaxy formation, e.g., the Illustris project (\citealt{2014MNRAS.444.1518V, 2014Natur.509..177V, 2014MNRAS.445..175G, 2015A&C....13...12N}), the Eagle project (\citealt{2015MNRAS.446..521S}); the Magneticum simulation (Dolag et al., in prep; \citealt{2018MNRAS.480.4636S}), the Horizon-AGN simulation (\citealt{2014MNRAS.444.1453D, 2014MNRAS.445L..46W, 2015MNRAS.452.2845K}), we are now equipped with a powerful numerical tool to address some of these fundamental and open issues. In particular, the latest cosmological magneto-hydrodynamical simulation -- the IllustrisTNG simulation (TNG hereafter; \citealt{2018MNRAS.475..676S, 2018MNRAS.475..648P, 2018MNRAS.475..624N, 2018MNRAS.480.5113M, 2018MNRAS.477.1206N}) has successfully reproduced many observational galaxy properties and scaling relations. For example, the observed mass$-$size relations in both late-type and early-type galaxies have been well recovered within observational uncertainties (\citealt{2018MNRAS.474.3976G, 2019MNRAS.483.4140R, 2019arXiv190307625H}). The simulation has also well reproduced the observed galaxy color bimodality distribution and mass$-$color relation (\citealt{2018MNRAS.475..624N}). Morphological and shape parameters as well as their mass dependencies of the TNG galaxies have also been studied in great detail and shown to be within $1\sigma$ scatter of the observed trends (\citealt{2019MNRAS.483.4140R}).

Despite their many successes, the most advanced cosmological simulations still disagree with observations in multiple aspects and at different degrees, which suggests that there may be missing links and/or incorrect ingredients in the galaxy formation models  currently implemented in simulations. Moreover, detailed studies using these simulations have also raised further puzzles from a theoretical point of view. To this end, stellar orbit de-composition serves a complementary role. It is not a property that can be tuned in a simulation in order to match observations, but an emergent phenomenon during galaxy formation and evolution. It can therefore help us to identify drawbacks in existing galaxy formation models, and to tackle the puzzles in combination with studies of other fundamental galaxy properties.

This work aims to present a first study of the stellar orbit compositions of the TNG galaxies and their relations with other fundamental galaxy properties, including stellar mass, size, color, morphology and star-formation rate, in different types of galaxies. In particular, we compare our results with an observational sample that is composed of 260 CALIFA (Calar Alto Legacy Integral Field Area) galaxies from \citet{2018NatAs...2..233Z} (Zhu18 hereafter), where an orbit-based Schwarzschild modelling technique (\citealt{1979ApJ...232..236S, 1993ApJ...409..563S}) was applied in order to derive a galaxy's stellar orbit composition. This study provided the first application of this sophisticated method to such a large number of observed galaxies, enabling for the first time statistical analyses and comparisons over wide ranges of galaxy mass and morphology.

The paper is organized as follows. In Sect.\,2, we introduce both the observed and the simulated galaxy samples. In Sect.\,3, we describe the method that we use to calculate stellar orbit compositions for both samples. In Sect.\,4, we present results of the simulated orbital distributions, their comparisons to those of the observed galaxies (Sect.\,4.1), and their relations to other fundamental galaxy properties (Sect.\,4.2). Conclusions and discussion are given in Sect.\,5. In this work, we adopt the Planck cosmology (\citealt{2016A&A...594A..13P}) in a flat universe, i.e., a total matter density of $\Omega_{\rm m}$ = 0.3089 (with a baryonic density of $\Omega_{\rm b}$ = 0.0486), a cosmological constant of $\Omega_{\Lambda}$ = 0.6911, and a Hubble constant $h=H_0/(100\kms\Mpc^{-1})=0.6774$, also used in the IllustrisTNG simulation.

\section{Observed and simulated samples}

\subsection{CALIFA kinematic sample}

The comparison sample is composed of 260 nearby galaxies from the CALIFA-kinematic galaxy sample (\citealt{2017A&A...597A..48F}), which was drawn from the CALIFA mother and extended samples (\citealt{2012A&A...538A...8S, 2016A&A...594A..36S}).

Selected from the photometric catalogue of the SDSS DR7 (\citealt{2009ApJS..182..543A}), a sample of 600 local-Universe ($0.005<z<0.03$) galaxies that span a variety of morphological types were targeted by the CALIFA survey for their spatially resolved spectroscopic information. Owing to the well-defined selection function within the stellar mass range between $10^{9.7}M_{\odot}$ to $10^{11.4}M_{\odot}$, the CALIFA galaxy sample can be volume-corrected to represent averaged properties of present-day galaxies (see \citealt{2014A&A...569A...1W} for details).

The CALIFA-kinematic sample was selected to include 300 galaxies with high-quality kinematic maps and without indications of being disturbed by nearby companions or recent merger activities. A Kolmogorov-Smirnov test was adopted to confirm that the stellar mass functions of the kinematic and the mother samples are statistically consistent (see \citealt{2017A&A...597A..48F}).

In Zhu18, an orbit-based Schwarzschild modelling technique was applied to the 260 galaxies in the CALIFA-kinematic sample. The resulting stellar circularity distributions exhibit a clear kinematic ``Hubble sequence'', where lower-mass ($\Mstar \la 10^{11} M_{\odot}$) galaxies are largely composed of kinematically warm and cold (ordered rotation) orbits, in contrast to galaxies at the high-mass end which are dominated by kinematically hot orbits (``random'' motion). This was the first time that such a sophisticated orbit-based modelling method was ever applied to such a large number of observed galaxies, enabling statistical analyses over wide ranges of galaxy mass and morphology. For this reason, we use these data as our comparison sample.

\subsection{TNG galaxy sample and selection criteria}

The simulated galaxies that we select for the study are taken from the TNG100 simulation (\citealt{2018MNRAS.475..676S, 2018MNRAS.475..648P, 2018MNRAS.475..624N, 2018MNRAS.480.5113M, 2018MNRAS.477.1206N}) -- a magnetohydrodynamic cosmological simulation, computed with $2 \times 1820^3$ resolution elements in a cosmological box of $\sim(110\Mpc)^3$ (reaching a baryonic resolution of $\sim 1.4 \times 10^6M_{\odot}$ and a spatial resolution of $\sim 0.74\kpc$), using the advanced moving-mesh code {\sc arepo} (\citealt{2010MNRAS.401..791S}). The TNG100 simulation is currently publicly available at http://www.tng-project.org (see \citealt{2019ComAC...6....2N}). Based on the original Illustris models (\citealt{2013MNRAS.436.3031V, 2014MNRAS.438.1985T}), the TNG simulation has adopted new physics models and improved implementations of galactic winds, stellar evolution, chemical enrichment (\citealt{2018MNRAS.473.4077P}), as well as AGN feedback (\citealt{2017MNRAS.465.3291W}), such that the simulation has better reproduced many observed galaxy properties and scaling relations to different degrees, which we do not review here in detail.

Galaxies in their host dark-matter halos were identified using the {\sc subfind} halo finding algorithm (\citealt{2001MNRAS.328..726S, 2009MNRAS.399..497D}). The total halo mass, $\Mhalo$, is defined as the mass from all simulation particles/elements that are gravitationally bound to the galaxy. The $3D$ half-stellar-mass radius $R_{*}^{\rm h3 \it D}$ of a galaxy is defined as the radius that encloses half of the total stellar mass bound to the host halo of this galaxy. Both quantities above are obtained from the simulation's {\sc subfind} catalogue. In addition, for each galaxy we explicitly define a stellar mass, $\Mstar$, and a SDSS $r$-band magnitude, $M_r$, that are associated with the tightly bound stellar component at the centre of the host dark-matter halo. They are calculated using all stellar particles within a $3D$ radius of 30 kpc from the galaxy centre. This radius has been adopted to exclude contributions from halo stars or intra-cluster light for the sake of fair comparisons to galaxy observations (e.g., \citealt{2015MNRAS.446..521S, 2018MNRAS.475..648P}).

%{\it Still need to look at Enviroment! [Yuan Wang]} 
Regarding sample selection, galaxies from the CALIFA-kinematic sample are truly a mixed bag in the local Universe: they have a variety of morphologies, span a wide range in stellar mass, and live in different (non-cluster) environments. In order to form a simulated sample that can be compared to the observed galaxies, we randomly selected, from the $z=0$ snapshot, one out of five galaxies that have stellar masses $\Mstar/M_{\odot}\in[10^{9.7},\,10^{11.4}]$ and total halo masses $\Mhalo/M_{\odot}\in[10^{11}, \,10^{14}]$, regardless of galaxy type, or of being a central galaxy or a satellite. The mass ranges used above were motivated by the CALIFA-kinematic galaxy observation (see \citealt{2014A&A...569A...1W} for details). These selection criteria result in $\sim$ 1500 galaxies (among which 74\% (26\%) are central (satellite) galaxies); the smallest galaxies in our sample are resolved by more than 3500 stellar particles within 30 kpc from the galaxy centres.

Table 1 presents ranges of fundamental galaxy/halo properties listed above for both observation and simulation samples. Note that we specifically choose to use $R_{*}^{\rm h3 \it D}$ for the simulated galaxies, in comparison to the half-light semi-major axes, denoted by $R_{\rm eff}^{\rm maj}$, of the CALIFA-kinematic galaxies, although the two definitions are not the same technically. As is described in Sect.\,3.1, the observed fractions of different orbital components are evaluated within $R_{\rm eff}^{\rm maj}$. To make fair comparisons, the orbital evaluation for the simulated galaxies shall also be carried out within similar radial ranges. As can be seen in Table 1, the ranges of $R_{*}^{\rm h3 \it D}$ for the simulated galaxies and $R_{\rm eff}^{\rm maj}$ for the observed galaxies are similar. Therefore, instead of calculating and adopting the strictly defined $R_{\rm eff}^{\rm maj}$ for the simulated galaxies, we simply use $R_{*}^{\rm h3 \it D}$ as the evaluation aperture radius for a given galaxy. We note that the observed galaxy mass$-$size relations have been well reproduced by the simulation within observational uncertainties (\citealt{2018MNRAS.474.3976G, 2019MNRAS.483.4140R, 2019arXiv190307625H}).

\begin{table}
\caption{Ranges of SDSS $r$-band magnitude, $M_r$, and evaluation aperture radius, $R_*$, for both observation and simulation samples, where galaxies satisfy $\Mstar/M_{\odot}\in[10^{9.7},\,10^{11.4}]$ and $\Mhalo/M_{\odot}\in[10^{11},\,10^{14}]$ (see \citealt{2014A&A...569A...1W}). }
\begin{minipage}{\textwidth}
\begin{tabular}{l | c | c | c} \hline
Sample~~~~~~~~~~ & ~~~~~$M_r$~~~~~ & ~~~~~$R_* [\kpc]$~~~~~  \\\hline\hline
CALIFA-kinematic & [-23.1,\,-19.0]~~ & ~~[1.4,\,22.7]~ ($R_{\rm eff}^{\rm maj}$)~~ \\\hline
TNG100 galaxies & [-23.1,\,-18.6]~~ & ~~[1.4,\,27.5]~ ($R_{*}^{\rm h3 \it D}$)~~ \\
\hline\hline
\end{tabular}
\end{minipage}
\label{tab:Table1}
\end{table}

\section{Methodology}

In this paper, we calculate fractions of different stellar orbital components sampled by stars (stellar particles), whose circularities, $\lambda_{z}$, are used to classify different orbital types. For a given star with a binding energy $E$ ($<0$), $\lambda_{z}$ is defined as the ratio between $L_{z}$ and $J_{\rm c}(E)$, where $L_{z}\equiv xv_y-yv_x$ is the magnitude of the angular momentum component along the shortest axis $z$ (closest to the spin axis) of the galaxy; and $J_{\rm c}(E)$ is the maximum $L_{z}$ (corresponding to the circular orbit) among all stars that have a binding energy of $E$ (see Sect.\,3.1 and 3.2 for a detailed description).

Generally speaking, stars on circular orbits that rotate about the same direction as the bulk angular momentum ${\vc L}_{\rm gal}$ always have $\lambda_{z}\approx 1$; they are the constituents that form the flat and co-rotating disk feature of a galaxy. Stars on radial or boxy orbits (which can come arbitrarily close to the galactic center) have $\lambda_{z}\approx 0$ on average over several orbital periods; they are the constituents that form the more spherical-shaped bulge component, whose dynamical state is often described as ``random'' motion in a statistical sense. There are also stars which consistently have negative $\lambda_{z}$, rotating about the opposite direction as ${\vc L}_{\rm gal}$. Theoretically, these fundamentally different orbits are expected to have different formation origins (e.g., \citealt{2004ApJ...612..894B, 2010ApJ...723..818H, 2013ApJ...773...43B, 2013MNRAS.436..625S, 2016MNRAS.459..199G, 2017MNRAS.472.3722G, 2018MNRAS.480.4636S, 2019arXiv190303627S}).

For an observed galaxy, orbit-based dynamical modelling can readily derive information about {\it time-averaged} stellar circularities. This provides a basis for simple orbital family classification (e.g., Zhu18). From the simulation perspective, however, the {\it time-averaged} circularity is not cheap to compute when one has a large number of galaxies and each is sampled by a fairly large number of stellar particles. A major reason is that in order to conduct correct orbital integration, real-time potential and force re-evaluations have to be carried out with sufficiently small time steps, which significantly increases the computational cost.

A cheaper approach is to the calculate the {\it instantaneous} stellar circularity for orbital classification, which depends only on the particle's state and the overall gravitational potential at a given snapshot (thus without need for time integration). Caution has to be applied, however, that the adoption of {\it instantaneous} circularity distributions requires careful calibration using the true distribution, i.e., the {\it time-averaged} circularity distribution, which is what we have done in this work (see Sect.\,3.1). This is because the latter distribution tends to be ``spread out'' (in particular the peak features at around $\lambda_{z} \approx 0$ and 1) at any given instant of time, e.g., a star that possesses some very ``hot'' radial orbit with {\it average} circularity $\lambda_{z} \approx 0$ can have an {\it instantaneous} circularity $\lambda_{z}\gg 0$ at one time and $\lambda_{z} \ll 0$ at some other time. 

In Sect.\,3.1, we discuss how we calculate the {\it instantaneous} circularity distributions for the CALIFA-kinematic galaxies, and we also show that the relative fractions of four different orbit types that we classify using {\it instantaneous} circularities are statistically the same as those derived using {\it time-averaged} quantities. In Sect.\,3.2, we present the method that we adopt to calculate the orbit fractions for the simulated galaxy sample.

\subsection{Instantaneous circularity distribution and orbital types for the observational sample}

In Zhu18, the {\it time-averaged} circularity distribution of any given observed galaxy was calculated from the best-fit orbit-superposition model, which reproduced a galaxy's observed surface brightness, mean velocity and velocity dispersion maps. Here, we recount briefly the key steps in their work: a parameterized galaxy density/potential model was assumed, where a triaxial stellar distribution based on a deprojected $2D$ Multiple Gaussian Expansion model that best fit the galaxy image was embedded in a spherical NFW dark matter distribution; for any given parameterization, an orbital library made of $10^4$ orbits was first constructed by test particle integration under the given gravitational potential; for each orbit $i$, a weight $w_i$ was sought such that the combined solution would minimize the residual $\chi^2$ between model predictions and observational datasets on surface brightness and kinematic distributions. Note that $\Sigma_i w_i = 1$. The best-fit model corresponds to the smallest (and satisfactory) $\chi^2$ among all explored parameterizations. In order to calculate the {\it time-averaged} circularity distribution, each orbit (under the best-fit model) with binding energy $E$ could be associated with a time-averaged circularity $\overline{\lambda_z} \equiv \overline{L_z} / J_{\rm c}(E)$, where $\overline{L_z}\equiv \overline{x v_y - y v_x}$ was the time-averaged angular momentum along the $z$- (shortest principal) axis; $J_{\rm c}(E)$, as the largest $z$-angular momentum corresponding to the circular orbit of energy $E$, was approximated by $\overline{r} \times \overline{V_{\rm rms}}$, where $\overline{r}$ is the time-averaged $3D$ galacto-centric radius of this orbit and $\overline{V_{\rm rms}}$ the time-averaged second velocity moment $V_{\rm rms} \equiv \sqrt{(v_x^2+v_y^2+v_z^2  +2v_xv_y+2v_yv_z+2v_xv_z)}$. The final statistics of {\it  time-averaged} orbital circularity for each observed galaxy only used those orbits that had $\overline{r}\leqslant R_{\rm eff}^{\rm maj}$. We note that this best-fit galaxy density/potential model together with its orbital library \{$i$\} and orbital weights \{$w_i$\} are then our starting point to calculate the {\it instantaneous} orbital circularity distribution for a given observed galaxy.

To do so, we sample the stellar matter distribution of a given observed galaxy with a total of $10^7$ particles. Specifically, each orbit $i$, which has a weight of $w_i$ (in the best-fit model), is randomly (uniformly) sampled in time during 50 orbital revolutions using $10^7 w_i$ particles (i.e., the orbit is sampled randomly for $10^7 w_i$ times), whose position and velocity pair are given according to the exact time lapse into the orbital evolution where the particle is drawn. Galacto-centric distance, $r$, and instantaneous circularity, $\lambda_z \equiv L_z / J_{\rm c}(E)$, of each sampling particle can then be obtained. Note that for a given orbit, the sampled $L_z$ values can be different among different sampling particles; while their $J_{\rm c}$ values, as taken from the time-averaged calculation, are kept the same. Again, we only select particles with $r \leqslant R_{\rm eff}^{\rm maj}$ for the final statistics of the instantaneous circularity distribution.

To broadly classify instantaneous orbital types, we divide the sampling particles into four different components according to their $\lambda_z$ values: (1) a cold component is composed of particles with $\lambda_z\geqslant0.8$; (2) warm orbits correspond to particles with $0.25<\lambda_z<0.8$; (3) a hot component is composed of particles with $-0.25\leqslant\lambda_z\leqslant0.25$; and (4) finally counter-rotating orbits correspond to particles with $\lambda_z<-0.25$.

For each component, we then calculated its $r$-band luminosity fraction $f_{\lambda_z}(r\leqslant R_{\rm eff}^{\rm maj})$, which is defined as the ratio between the luminosity of all the constituent particles of this component and the total luminosity from all particles within the given radial range. The luminosity fractions for the cold, warm, hot and counter-rotating orbit components are denoted as luminosity fraction $f_{\rm cold}$, $f_{\rm warm}$, $f_{\rm hot}$ and $f_{\rm CR}$, respectively.

\begin{figure}
\centering
\includegraphics[width=8cm]{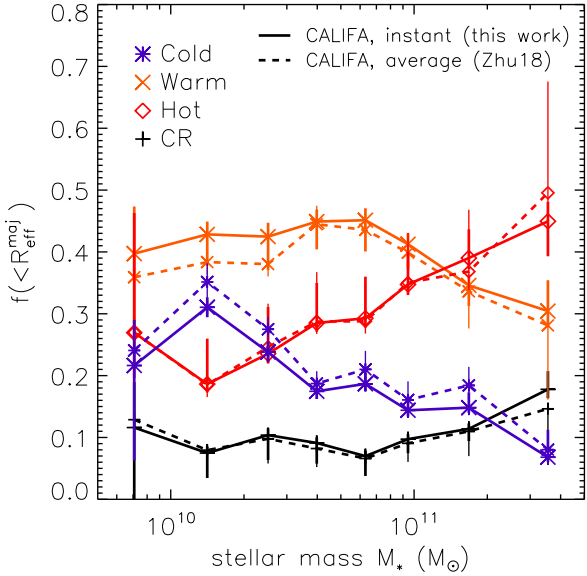} 
\caption{Volume-corrected average luminosity fractions of different orbital components as functions of stellar mass, for the 260 CALIFA galaxies in Zhu18. $f_{\lambda_z}(r\leqslant R_{\rm eff}^{\rm maj})$ of four different orbital types are given in different symbols. Those that are connected by the solid and the dashed lines indicate the averaged results using {\it instantaneous} (this work). and {\it time-averaged} (Zhu18) stellar circularities, respectively. The error bars indicate $1\sigma$ lower and upper uncertainties of the mean due to both statistical errors and model biases (for more details see on-line material of Zhu18). }
\label{fig:CALIFA2}
\end{figure}

Fig.\,\ref{fig:CALIFA2} shows the volume-corrected average luminosity fraction $f_{\lambda_z}(r\leqslant R_{\rm eff}^{\rm maj})$ for four different orbital types that are classified using both {\it instantaneous} (this work) and {\it time-averaged} (Zhu18) stellar orbit circularities, as functions of stellar mass, for the CALIFA galaxies in Zhu18. The error bars indicate $1\sigma$ lower and upper uncertainties of the mean due to both statistical errors and model biases; the latter are obtained by rigorous tests against simulations (for details, see on-line material of Zhu18). The means and uncertainties are also given in Table \ref{tab:CALIFA_combined}. As can be seen, a small fraction of stars ($<5\%$) that genuinely occupy cold orbits are now classified as possessing the warm orbits when using the {\it instantaneous} circularity distribution for orbital classification. This is seen across nearly the entire mass range investigated here. While in the most-massive bin, where the hot component dominates galaxy kinematics, a small portion of the kinematically hot stars ($<5\%$) are now contributing to the warm and counter-rotating orbit fractions. In spite of these small differences, it is fair to say that such a coarse classification (with relatively large ranges in $\lambda_{z}$) results in consistent orbital fractions (within model uncertainties) between using the two different circularity definitions. From now on, we use the result from {\it instantaneous} stellar circularities for comparisons with the simulation.

\begin{table*}
\caption{Volume-corrected average luminosity fractions (SDSS $r$-band) of cold, warm, hot and counter-rotating components in eight stellar mass bins (as plotted in Fig.\,\ref{fig:CALIFA2}) for the 260 CALIFA galaxies in Zhu18. The presented fractions are derived using {\it instantaneous} orbit circularities. For each component, $1\sigma$ uncertainties of the mean are also given, which take into account both model biases and statistical errors (for more details see  on-line material of Zhu18). }
\begin{minipage}{\textwidth}
\begin{tabular}{l | c | c | c | c | c | c | c | c } \hline
  $\log M_*$ range & [9.7, 10.0] & [10, 10.3] & [10.3, 10.5] & [10.5, 10.7] & [10.7, 10.9] & [10.9, 11.05] & [11.05, 11.4] & [11.4, 11.7]  \\\hline\hline
$f_{\rm cold}(r\leqslant R_{\rm eff}^{\rm maj})$ & 0.22 & 0.31 & 0.24 & 0.17 & 0.19 & 0.14 & 0.15 & 0.07 \\
Lower uncertainty &  -0.15 & -0.02 & -0.02 & -0.01 & -0.01 & -0.01 & -0.01 & -0.01 \\
Upper uncertainty & 0.07  &  0.02 &  0.02 &  0.02 &  0.02 &  0.02 & 0.02 & 0.04 \\\hline
$f_{\rm warm}(r\leqslant R_{\rm eff}^{\rm maj})$ & 0.40 & 0.43 & 0.42 & 0.45 & 0.45 & 0.41 & 0.35 & 0.30 \\
Lower uncertainty & -0.14 & -0.05 & -0.05 & -0.05 & -0.05 & -0.04 & -0.03 & -0.03 \\
Upper uncertainty & 0.08 &  0.02 &  0.02 &  0.02 &  0.02 &  0.02 &  0.02 & 0.05 \\\hline
$f_{\rm hot}(r\leqslant R_{\rm eff}^{\rm maj})$ & 0.27 &   0.19 &   0.24 &   0.29 &   0.29 & 0.35 &   0.39 &   0.45 \\
Lower uncertainty & -0.07 & -0.01 & -0.02 & -0.01 & -0.02 & -0.02 & -0.02 & -0.06 \\
Upper uncertainty  & 0.19 &  0.07 &  0.08 &  0.06 &  0.07 &  0.05 &  0.05  & 0.03 \\\hline
$f_{\rm CR}(r\leqslant R_{\rm eff}^{\rm maj})$  & 0.11 &  0.07 &   0.10 &  0.09 &  0.07 &  0.10 & 0.11 & 0.18 \\
Lower uncertainty & -0.13 & -0.04 & -0.04 & -0.03 & -0.03 & -0.02 & -0.02 & -0.01 \\
Upper uncertainty & 0.07 &  0.01 &  0.01 & 0.01 & 0.01 & 0.01 & 0.01 & 0.03  \\\hline\hline
\end{tabular}
\end{minipage}
\label{tab:CALIFA_combined}
\end{table*}

\subsection{Instantaneous circularity distribution and orbital types for the simulated sample}

For each simulated galaxy, we first identify the principal axes by finding the eigenvectors of the moment of inertia matrix that is associated with all stellar particles within a chosen $3D$ radius from the galaxy centre. This radius is set to be the smaller one between $3R_{*}^{\rm h3 \it D}$ and 30 kpc, in order to restrict the calculation to the visible galaxy region. We then rotate the galaxy so that the new $z$-axis is lined up with the shortest principal axis, whose positive direction is chosen to be the one that is closer to ${\vc L}_{\rm gal}$, which is the total stellar angular momentum of all stellar particles within the same radial range mentioned above.

In order to calculate the circularity distribution of a given galaxy, we randomly select a subsample of no more than 50,000 particles (to reduce computational cost) from all the stellar particles that are located within 30 kpc from the galaxy centre. For a TNG100 galaxy with $\Mstar/M_{\odot}\in[10^{9.7},\,10^{11.4}]$, its circularity distribution is then sampled by 5,000 to 50,000 stellar particles. We carry out a convergence test and verify that the final circularity distribution converges for 50,000 down to 5,000 stellar particles.

For each selected sampling particle, its instantaneous circularity $\lambda_z \equiv L_{z}/J_{\rm c}(E)$ can be readily calculated using its phase space information. In particular, its binding energy $E$ is calculated as the sum of its kinetic and gravitational potential energy. In order to calculate the corresponding $J_{\rm c}(E)$, we first rank all sampling particles by their $E$ values; and for a given particle, we then find the highest $|L_{z}|$ among its 100 neighbouring particles in $E$; this $|L_{z}|$ is then taken as $J_{\rm c}(E)$ for this particle (see \citealt{2014MNRAS.437.1750M, 2015ApJ...804L..40G}). We verify that $J_{\rm c}(E)$ defined in this way is consistent with the quantity defined in Sect.\,3.1 of Zhu18 for the CALIFA galaxies.

Following the same practice as described in Sect.\,3.1, we also divide the sampling stellar particles of each galaxy into four orbital types according to their instantaneous $\lambda_z$ values. For each component, we then calculate its $r$-band luminosity fraction $f_{\lambda_z}(r\leqslant R_{*}^{\rm h3 \it D})$ using sampling particles that are located within the galaxy's half-stellar-mass radius $R_{*}^{\rm h3 \it D}$.

\begin{figure} \centering
\includegraphics[width=4cm]{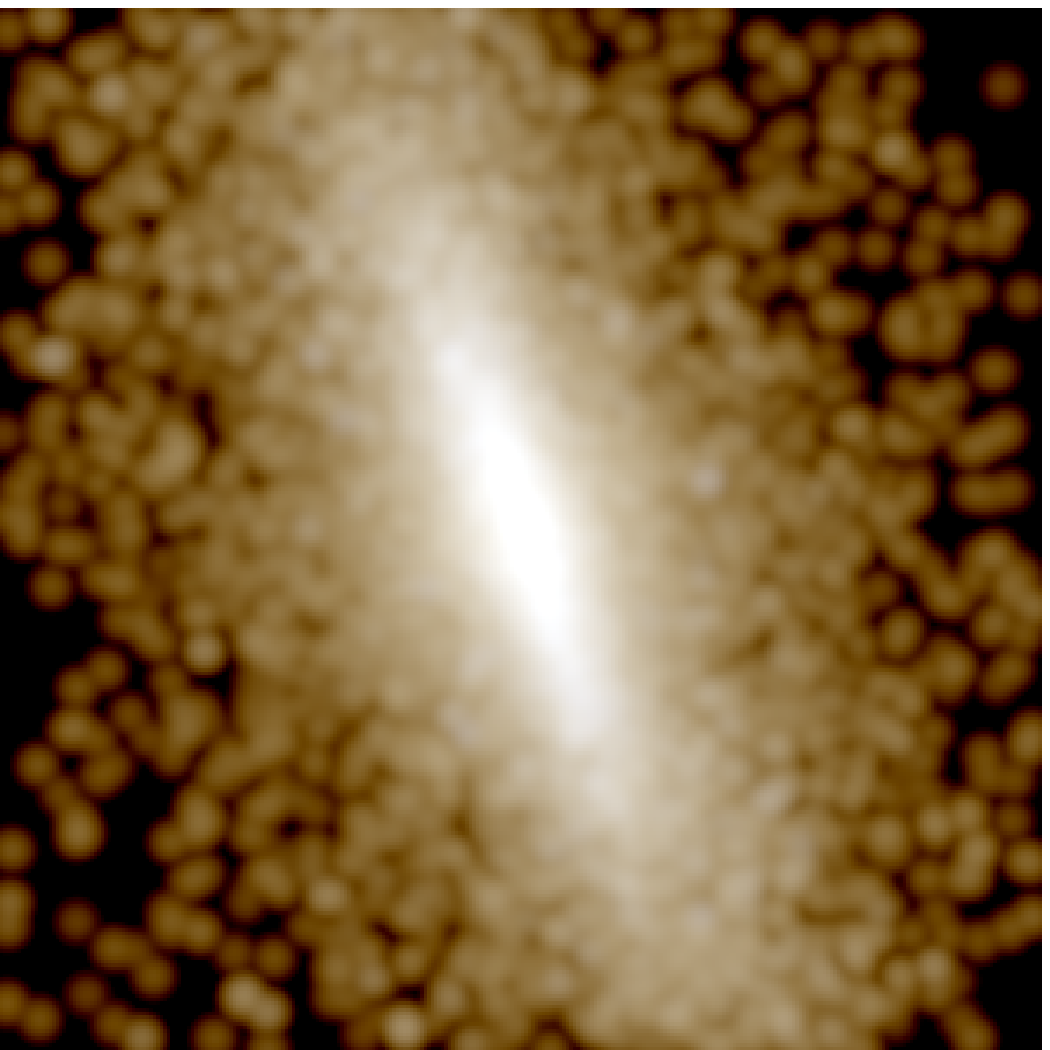}
\includegraphics[width=4cm]{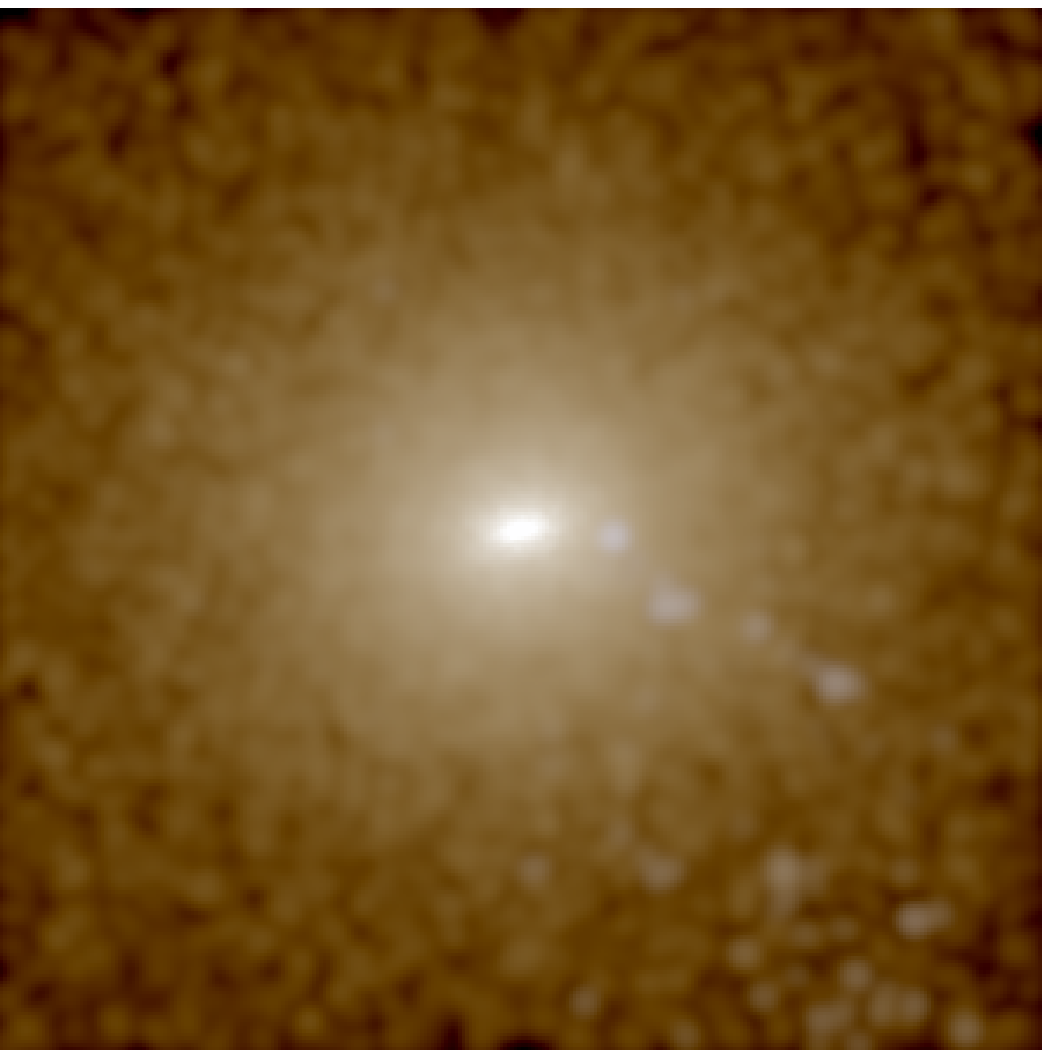}
\includegraphics[width=4cm]{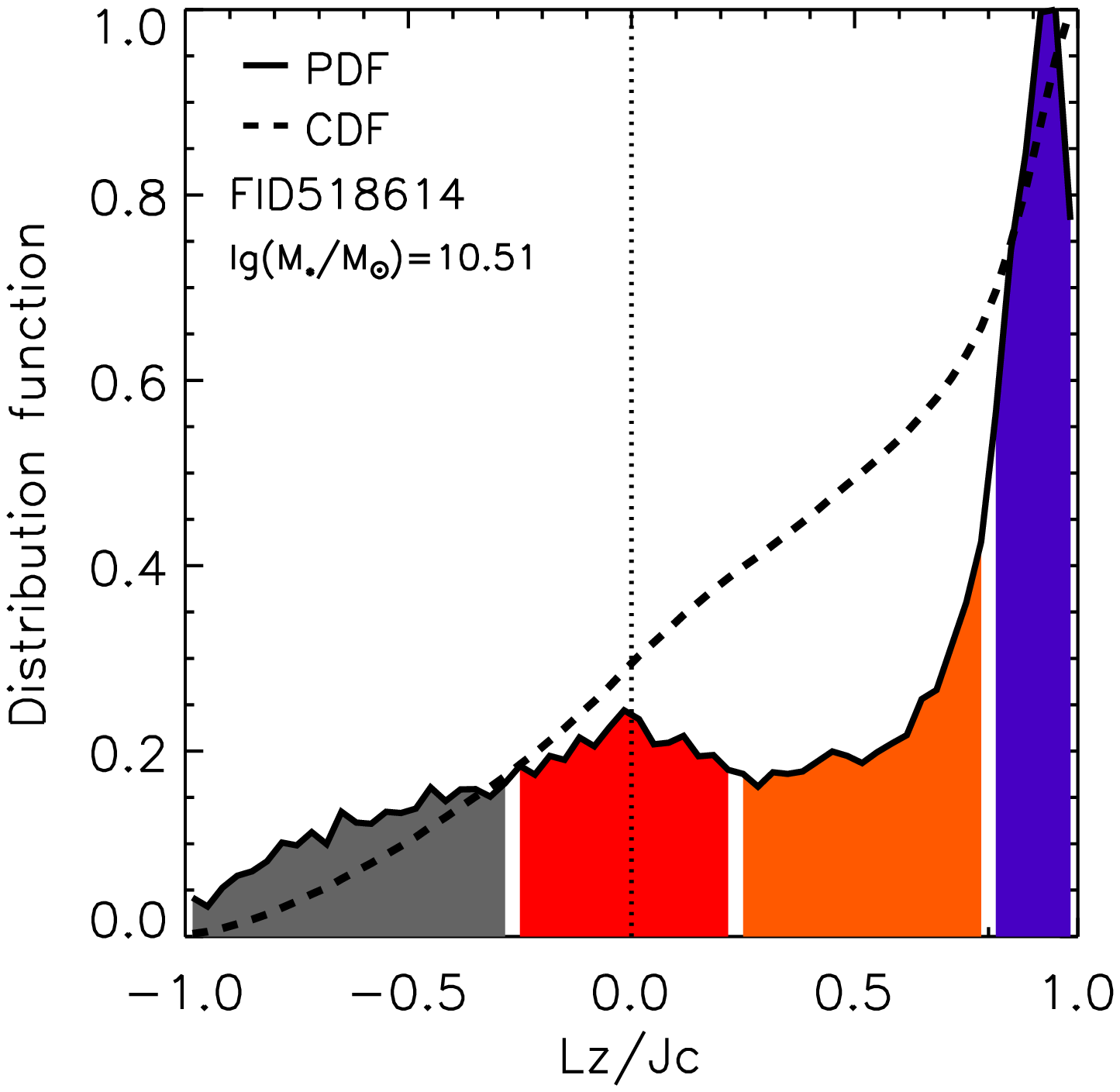}
\includegraphics[width=4cm]{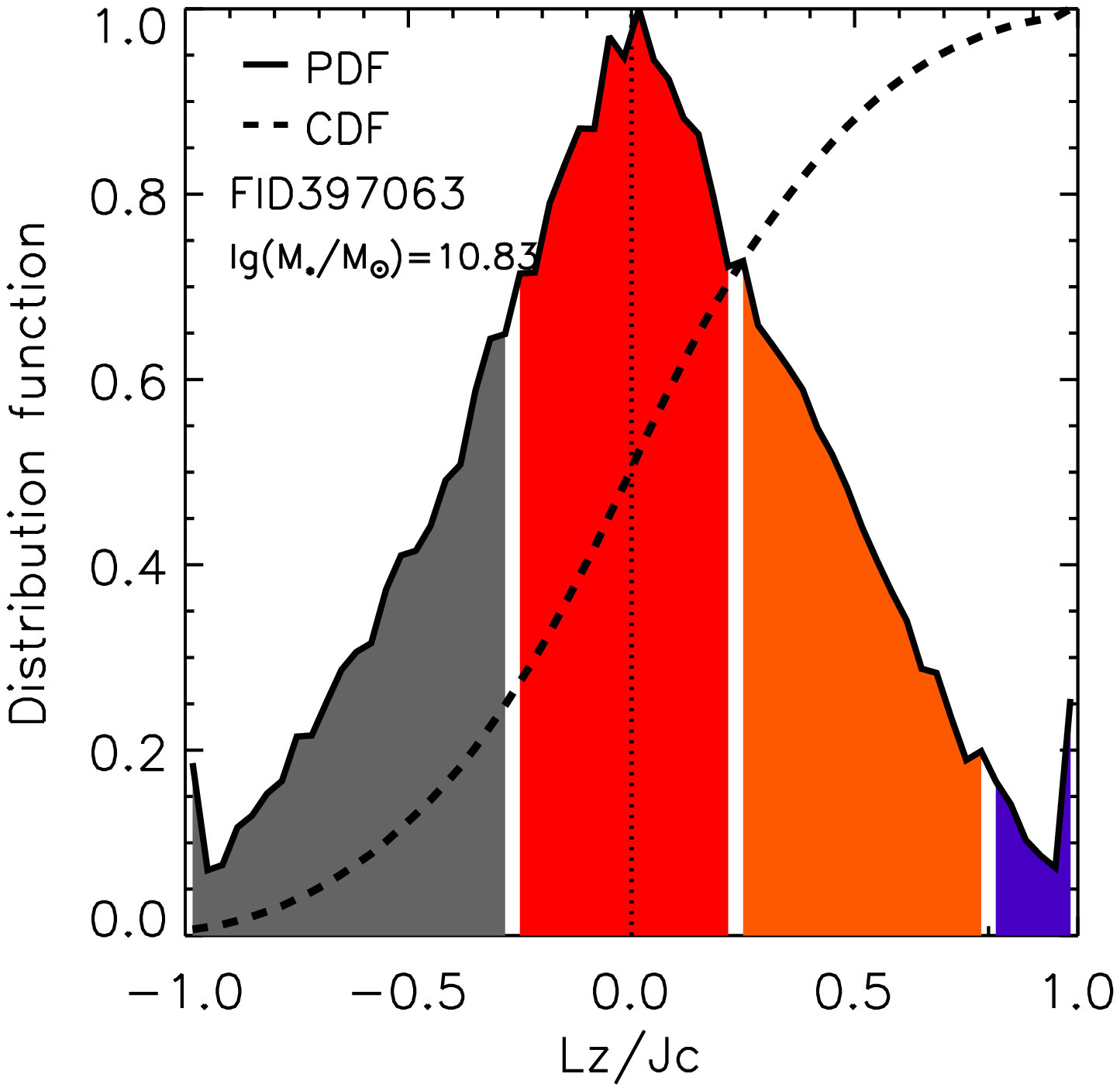} 
\caption{Top panels: an example disk galaxy with $(B/T)_{\rm photo}\sim 0.2$ and a S{\'e}rsic index of 1.0 (left) and an example elliptical galaxy with $(B/T)_{\rm photo} \sim 1.0$ and a S{\'e}rsic index of 4.8 (right). Bottom: the differential (solid) and cumulative (dashed) probability distribution functions of instantaneous stellar circularity, both of which are normalized by setting the peak values to be one, for the two galaxies above. The four colored regions under the cumulative distribution functions indicate, from left to right, the counter-rotating ($\lambda_z<-0.25$), hot ($-0.25\leqslant\lambda_z\leqslant0.25$), warm ($0.25<\lambda_z<0.8$) and cold ($\lambda_z\geqslant0.8$) components, respectively.}
\label{fig:LTET_Circularity} 
\end{figure}

We present in Fig.\,\ref{fig:LTET_Circularity} the differential (blue) and cumulative (red) probability distribution functions of instantaneous $\lambda_z$ for an example late-type galaxy, which has a bulge-to-total luminosity ratio of $(B/T)_{\rm photo}\sim 0.2$ and a S{\'e}rsic index of 1.0 (on the left), and an example early-type galaxy, which has $(B/T)_{\rm photo}\sim 1.0$ and a S{\'e}rsic index of 4.8 (on the right). To calculate $(B/T)_{\rm photo}$, a combined light profile model using a S{\'e}rsic (\citealt{1963BAAA....6...41S}) plus an exponential distribution is fitted to the $r$-band radial surface brightness distribution that is measured within elliptical isophotes; $(B/T)_{\rm photo}$ is then given by the ratio between the S{\'e}rsic and the total luminosities (see \citet{2017MNRAS.469.1824X} for more details). As can be seen, late-type galaxies in general have significant disk components with $\lambda_z$ peaking around 1; while early-type galaxies have well established bulge components with $\lambda_z$ peaking at about zero.

\subsection{Other galaxy properties and galaxy types}

To understand general inter-connections and dependencies on galaxy types, we calculate the following properties for the simulated galaxies: dust-attenuation corrected SDSS $(g-r)_{\rm c}$ color, SDSS $r$-band S{\'e}rsic index and bulge-to-total luminosity ratio $(B/T)_{\rm photo}$, and specific star formation rate (sSFR). The first three quantities are calculated using the method described in \citet{2017MNRAS.469.1824X}. The sSFR is defined as the ratio between a galaxy's star formation rate and stellar mass; the former is measured using all stellar particles that are projected within a radius of $R_{*}^{\rm h3 \it D}$ from the galaxy centre and that have star-forming ages of $\leq 1$ Gyr. We then explicitly define a late-type and an early-type galaxy population from the selected TNG100 galaxies, according to both their light-based morphologies and sSFR using a combination of three criteria:

\begin{itemize}
\item{\#1: A late-type galaxy is required to have $(B/T)_{\rm photo} < 0.5$, i.e., a smaller luminosity fraction in the galaxy bulge than in the disk is required. An early-type galaxy is required to meet the opposite criterion.}

\item{\#2: A late-type galaxy shall have its radial brightness distribution better fitted by an exponential profile than a de Vaucouleurs (\citealt{1948AnAp...11..247D}) profile within a radial range from 0.1 to $2\,R_{*}^{\rm h3 \it D}$. An early-type galaxy shall meet the opposite criterion.}

\item{\#3: Late-type and early-type galaxies can be separated using a certain sSFR threshold, which is a common practice in observations (\citealt{2011MNRAS.413..996M, 2013MNRAS.432..336W, 2014ApJ...782...33L, 2018PASJ...70S..23J}). Here we set the threshold to $-2$ for $\log[{\rm sSFR}/{\rm Gyr}^{-1}]$ (see \citealt{2019MNRAS.485.4817D} for a detailed study on the systematics of different criteria to define main-sequence and quenching galaxies). }

\end{itemize}

Galaxies that meet both morphological criteria \#1 and \#2 in order to be classified as either late-types or early-types, make up $\sim 80\%$ of the total sample; galaxies that simultaneously meet criteria \#1 and \#3 (or \#2 and \#3) take up $\sim 65\%$ of the total sample; while $\sim 60\%$ of all selected galaxies would be explicitly classified into a specific galaxy type when using all three criteria at the same time. We adopt this last combination to compose typical but not necessarily complete samples of late- and early-type galaxies. In particular, this combination explicitly excludes a blue compact spheroidal population at lower masses and a red disk population at higher masses, which are produced by the simulation but are absent in the Pan-STARRS observations (see Figure 8 and 9 in \citealt{2019MNRAS.483.4140R}). 

We note that the comparisons to the observational data that are presented in Sect.\,4.1 are carried out using all selected galaxies in our sample regardless of their types. Only in Sect.\,4.2, we present the orbital fractions in relation to the above-stated fundamental properties in galaxies classified into different types according to the criteria above.

\section{Results}

\subsection{Comparisons between the TNG100 and CALIFA-kinematic samples}

\begin{table*}
\caption{Distributions of the luminosity fractions of the cold, warm,
  hot and counter-rotating orbits for the TNG100 galaxies in eight
  stellar mass bins (as plotted in Fig.\,\ref{fig:TNGvsCALIFA}). The
  presented average fractions are stellar-mass weighted, derived using
  {\it instantaneous} orbit circularities. Also given are the 5{\it
    th}, 16{\it th}, 84{\it th} and 95{\it th} percentiles of the
  distributions.}
\begin{minipage}{\textwidth}
\begin{tabular}{l | c | c | c | c | c | c | c | c } \hline
  $\log M_*$ range & [9.7, 10.0] & [10, 10.3] & [10.3, 10.5] & [10.5, 10.7] & [10.7, 10.9] & [10.9, 11.05] &\
 [11.05, 11.4] & [11.4, 11.7]  \\\hline\hline
$f_{\rm cold}(r\leqslant R_*^{{\rm h}3D})$ & 0.34 & 0.38 & 0.35 & 0.29 & 0.21 & 0.17 & 0.15 &  0.08 \\
5{\it th} percentile & 0.06 & 0.09 & 0.06 & 0.05 & 0.03 & 0.02 & 0.03 & 0.02 \\
16{\it th} percentile & 0.14 & 0.19 & 0.12 & 0.08 & 0.07 & 0.04 & 0.04 &  0.03 \\
84{\it th} percentile  & 0.53 & 0.58 & 0.56 & 0.57 & 0.33 & 0.31  & 0.29 & 0.11 \\
95{\it th} percentile  & 0.64 & 0.67 & 0.65 & 0.65  & 0.54 & 0.40 & 0.47 & 0.14 \\ \hline
$f_{\rm warm}(r\leqslant R_*^{{\rm h}3D})$ & 0.40 & 0.38 & 0.39 & 0.39 & 0.42 & 0.36 & 0.34 & 0.30 \\
5{\it th} percentile & 0.21 & 0.20 & 0.22 & 0.20 & 0.25 & 0.25 & 0.20 & 0.24 \\
16{\it th} percentile & 0.28 & 0.26 & 0.27 & 0.27 & 0.31 & 0.26 & 0.25 & 0.25 \\
84{\it th} percentile  & 0.50 & 0.50 & 0.52  & 0.49 & 0.51 & 0.45 & 0.44 & 0.38 \\
95{\it th} percentile  & 0.56 & 0.58 & 0.56 & 0.57 & 0.55 & 0.49 & 0.48 & 0.38 \\ \hline
$f_{\rm hot}(r\leqslant R_*^{{\rm h}3D})$ & 0.20 & 0.17 & 0.18 & 0.23 & 0.26 &  0.32 & 0.34 & 0.39 \\
5{\it th} percentile & 0.07 & 0.06 & 0.06 & 0.06 & 0.10 & 0.14 &  0.16  & 0.29 \\
16{\it th} percentile  & 0.10 & 0.09 & 0.09 & 0.10 & 0.14 & 0.20 & 0.22 & 0.31 \\
84{\it th} percentile  & 0.29 & 0.25 & 0.29 & 0.36 & 0.38 & 0.42 & 0.43 & 0.44 \\
95{\it th} percentile  & 0.39 & 0.34 & 0.37 & 0.41  & 0.44  & 0.48 & 0.45 & 0.46 \\ \hline
$f_{\rm CR}(r\leqslant R_*^{{\rm h}3D})$  & 0.07 & 0.07 & 0.08 & 0.10 & 0.10 & 0.15 & 0.17 & 0.22 \\
5{\it th} percentile & 0.02 & 0.01 & 0.01 & 0.01 & 0.02 & 0.03 & 0.04 & 0.16 \\
16{\it th} percentile  & 0.02 & 0.02 & 0.02 & 0.02 & 0.03 & 0.06 & 0.06 & 0.17 \\
84{\it th} percentile  &  0.11 & 0.09 & 0.13 & 0.16 & 0.18 & 0.25 & 0.27 & 0.25 \\
95{\it th} percentile  &  0.21 & 0.21 & 0.25 & 0.25 & 0.27 & 0.28 & 0.28  & 0.27  \\\hline\hline
\end{tabular}
\end{minipage}
\label{tab:TNG_distribution}
\end{table*}

\begin{figure*}
\centering
\includegraphics[width=8cm]{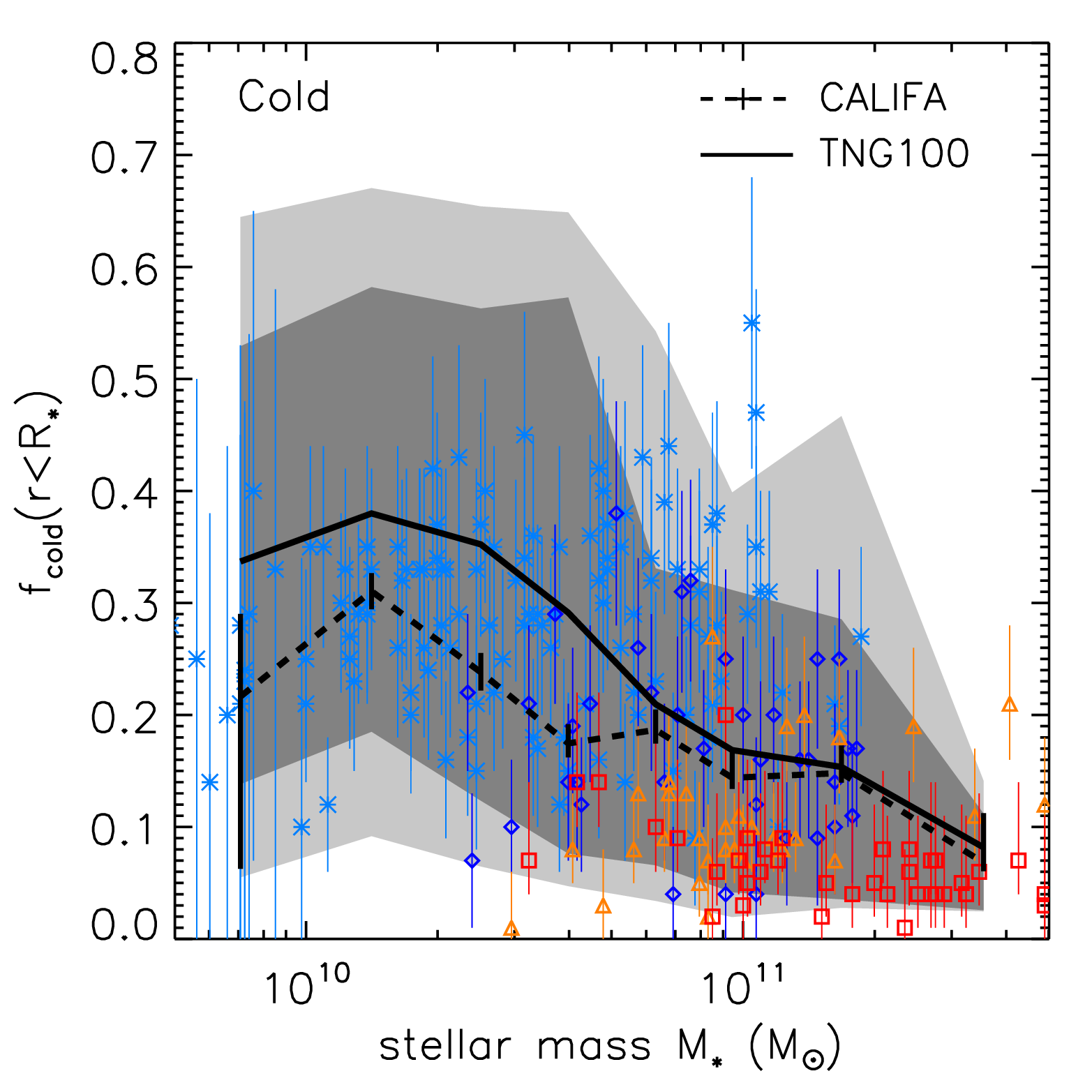}
\includegraphics[width=8cm]{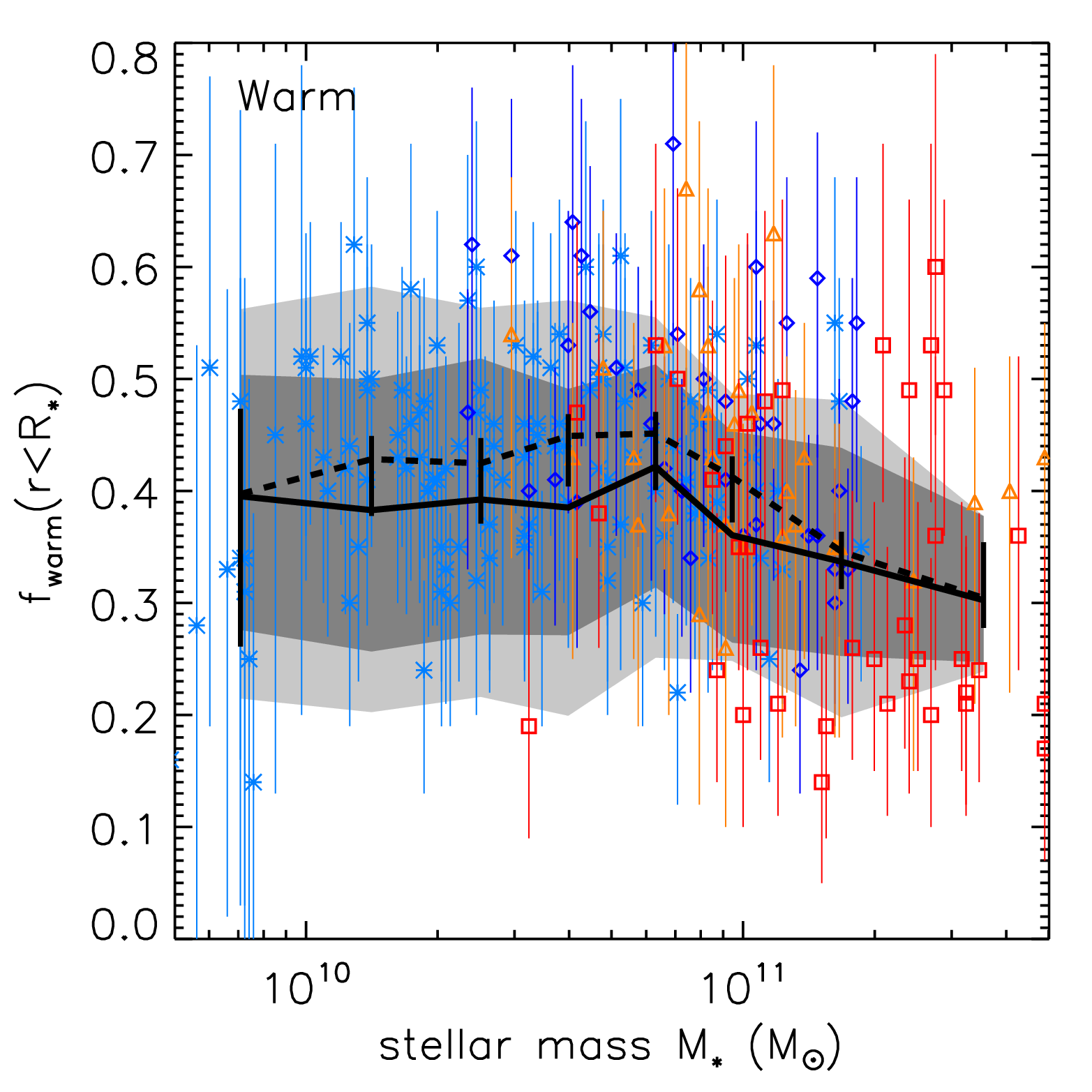} \\
\includegraphics[width=8cm]{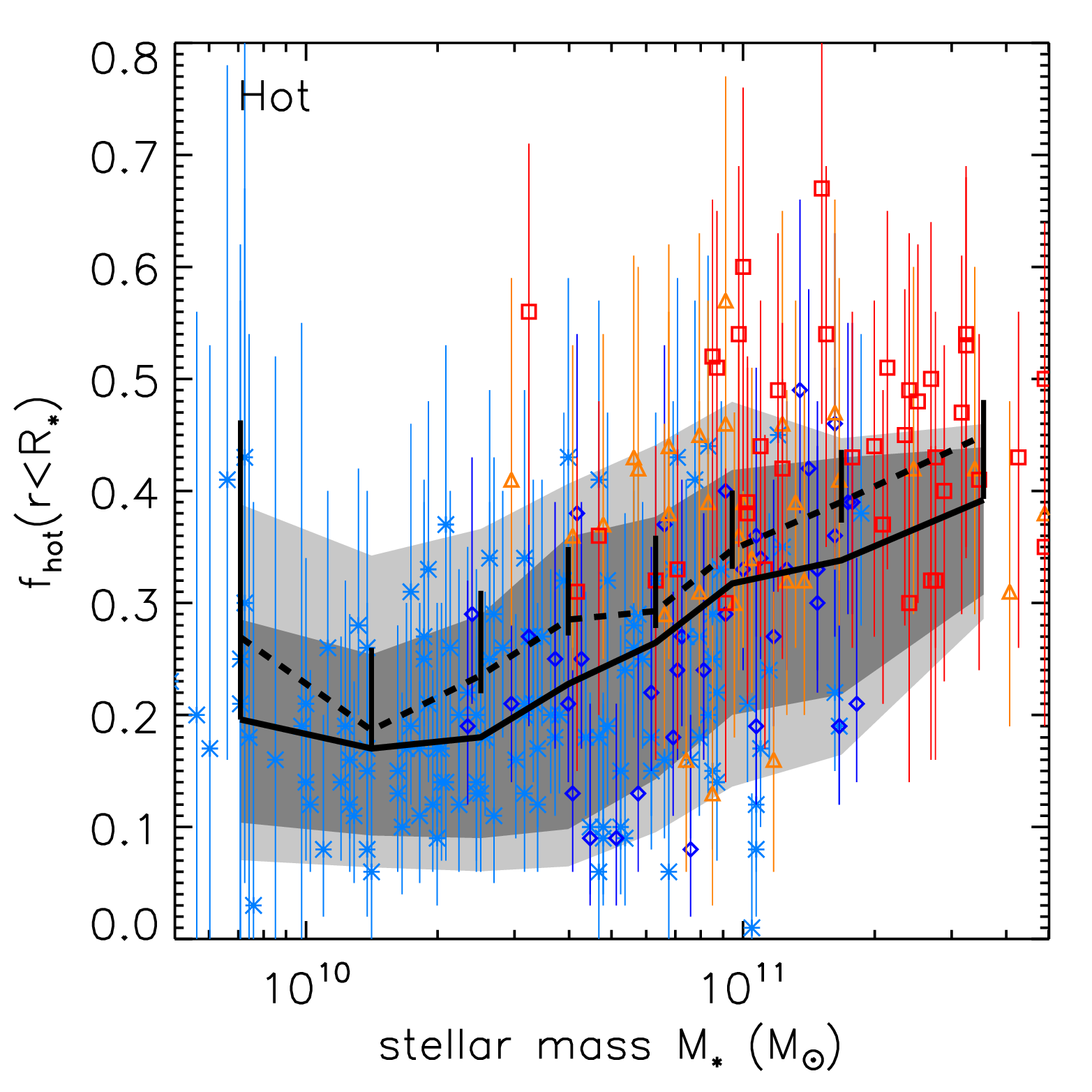}
\includegraphics[width=8cm]{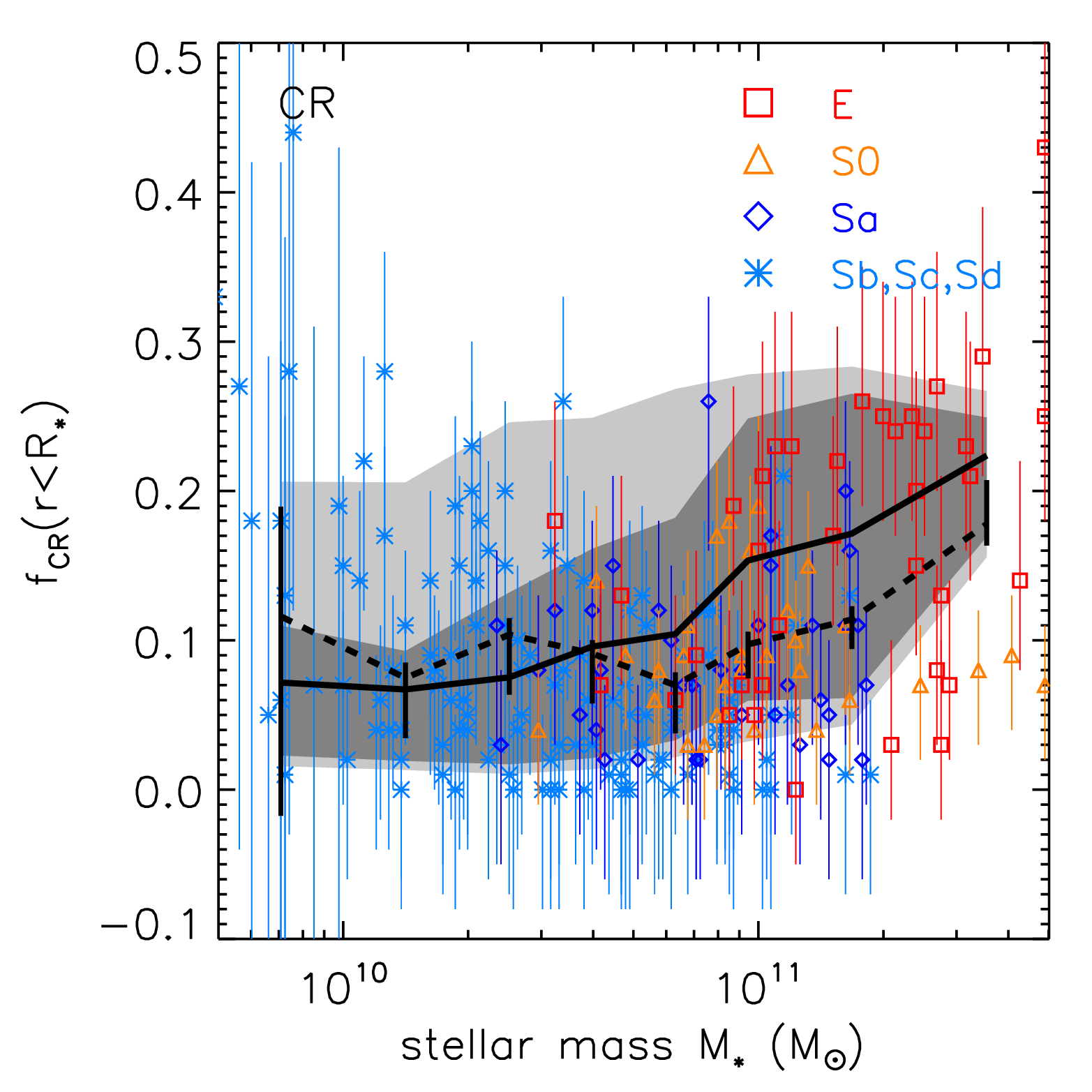} \\
\caption{Luminosity fractions $f_{\lambda_z}(r\leqslant R_*)$ of different orbital components, as functions of galaxy stellar mass $\Mstar$. Squares, triangles, diamonds and asterisks represent the E, S0, Sa, Sb$-$d galaxies, respectively, from the CALIFA-kinematic sample (Zhu18); error bars indicate systematic biases and uncertainties of the derived orbital fractions for individual systems. The dashed black lines mark the volume-corrected average luminosity fractions of the observational sample; the associated error bars indicate the upper and lower uncertainties of the means (same as Fig.\,\ref{fig:CALIFA2}). The solid black curves show galaxy-mass $\Mstar$-weighted average distributions of the TNG100 galaxy sample, whose 68\% and 90\% distributions around the median at any given mass are indicated by the grey and light grey shaded areas, respectively.}
\label{fig:TNGvsCALIFA}
\end{figure*}

We present in Fig.\,\ref{fig:TNGvsCALIFA}, luminosity fractions $f_{\lambda_z}(r\leqslant R_*^{{\rm h}3D})$ of the cold, warm, hot and counter-rotating orbit components, as functions of galaxy stellar mass $\Mstar$, for both the CALIFA and the full TNG100 galaxy samples. The means, 5{\it th}, 16{\it th}, 84{\it th} and 95{\it th} percentiles of the distributions for TNG100 galaxies are also given in Table \ref{tab:TNG_distribution}.

As can be seen, the TNG100 simulation broadly reproduces the volume-corrected average distributions of orbital luminosity fractions as functions of stellar mass, observed for the CALIFA-kinematic galaxies. In particular, the observed {\it mean} distributions of $f_{\rm warm}-M_{*}$ and $f_{\rm hot}-M_{*}$ are well recovered within modelling uncertainties of observed galaxies. Equally-well reproduced are the averaged $f_{\rm cold}-\Mstar$ and  $f_{\rm CR}-\Mstar$ relations at above and below $\Mstar \approx 6\times 10^{10}M_{\odot}$, respectively. In addition, the observed peak and trough features at $\Mstar \approx 1-2\times 10^{10}M_{\odot}$ in the mean distributions of $f_{\rm cold}-\Mstar$ and $f_{\rm hot}-\Mstar$, respectively, are broadly reproduced; above and below such a mass scale, the significance of the cold (hot) orbit component starts to diminish (grow) towards higher and lower massive systems, respectively, consistent with the kinematic Hubble Sequence as shown by the CALIFA galaxies in Zhu18. We note that \citet{2019arXiv190412860T} revealed the same sequence of the TNG100 galaxies encoded by the kinematic spheroid-to-total ratio and the stellar morphology concentration.

Noticeable disagreements are seen for the cold and counter-rotating components: below $\Mstar \approx 6\times 10^{10}M_{\odot}$ the simulated mean $f_{\rm cold}$ is higher than the observation by $\la 10\%$ (absolute orbital fraction); above this mass scale the simulated mean $f_{\rm CR}$ is higher than the observation also by $\la 10\%$ (absolute orbital fraction). Interestingly, even when allowing for observational uncertainties, it is still noticeable that the simulation produces much wider distributions of $f_{\rm cold}$, especially for galaxies of lower masses, and seemingly insufficient scatter at various mass scales for other orbital components.

\begin{figure*}
\centering
\includegraphics[width=16cm]{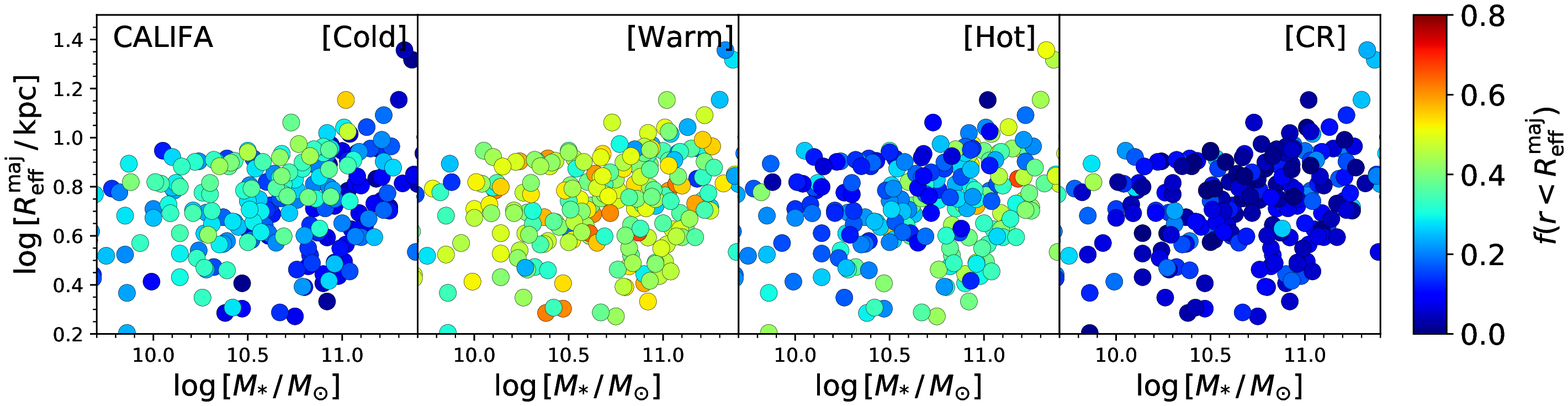}
\includegraphics[width=16cm]{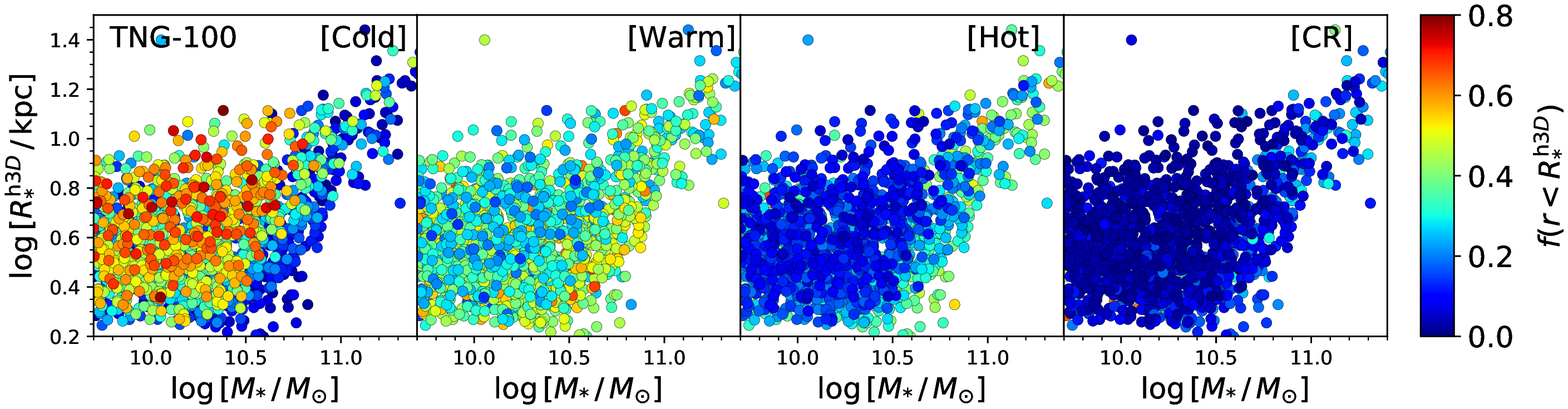}
\caption{The orbit-fraction evaluation aperture radius as a function of stellar mass for the CALIFA (upper panel) and the TNG100 (lower panel) galaxy samples, with $R_{\rm eff}^{\rm maj}-\Mstar$ plotted for the former and $R_{*}^{\rm h3 \it D}-\Mstar$ for the latter. The orbit luminosity fractions calculated in this work are evaluated within the two apertures for the two samples, respectively. Each circle represents a galaxy, color-coded by a given orbit luminosity fraction. Note that when plotting, galaxies in {\it all} panels are ranked by their $f_{\rm cold}$ such that higher-$f_{\rm cold}$ galaxies appear on top of lower-$f_{\rm cold}$ ones.}
\label{fig:SizeMass}
\end{figure*}

We consider whether the differences seen between the simulated and the observational mean distributions could simply be a result of unmatched evaluation apertures, as typically more ordered motion exists in the outskirts of a galaxy than at the centre and thus the larger the apertures are the higher the cold fractions are. We argue that this is not likely to be the major cause. On the one hand, the ranges of the evaluation aperture radii at a given stellar mass are very similar between the TNG100 and CALIFA samples, although the former has a small number of slightly larger-aperture outliers in comparison to the latter. This can be seen from Fig.\,\ref{fig:SizeMass}, where the evaluation-aperture radius is plotted as a function of stellar mass; each galaxy is color-coded by the measured orbit luminosity fraction (of a given type). Galaxies with overpredicted $f_{\rm cold}$ (left-most column) clearly exist in regions occupied by observational data.

On the other hand, we also adopt a different definition of ``effective radius'' to evaluate orbital fractions. In this case, the calculations are done within a radius that encloses half of the (SDSS $r$-band) luminosity projected within 30 kpc from the centre of a simulated galaxy. We confirm that this does not mitigate the noticeable difference seen between the simulated and the observational samples. However, we note that if the simulated galaxies were systematically smaller (while maintaining the overall range) and accordingly were to have systematically lower cold-orbit fractions, this could bring the simulation into closer agreement with observational data (but see Figure 7 of \citealt{2019MNRAS.483.4140R}, which shows the distribution of the TNG100 effective radii is well consistent with that of the Pan-STARRS observations).

In Sect.\,5, we discuss possible origins for the differences seen between the simulated and the observational orbital fractions from a perspective of the physics models adopted by the simulation. We also mention in passing that the overall distributions seen in Figs.\,\ref{fig:TNGvsCALIFA} and \ref{fig:SizeMass} are largely dominated by those of central galaxies (which take up $74\%$ of all selected sample). Satellite galaxies share qualitatively similar trends in terms of magnitude and scatter, albeit the exact distributions may differ. We leave the detailed differences and their connections with galaxy environment and accretion history to a future work, and from now on focus only on the central population of the simulated galaxies.

\subsection{Relations to other fundamental galaxy properties in different types of galaxies}

As an emergent phenomenon during the process of galaxy formation and evolution, a galaxy's kinematic status and orbit composition are entwined with other fundamental galaxy properties, such as mass, size, color, morphology and star formation rate. These fundamental properties vary with and thus also define different types of galaxies.

\begin{figure*}
\centering
\includegraphics[width=16cm]{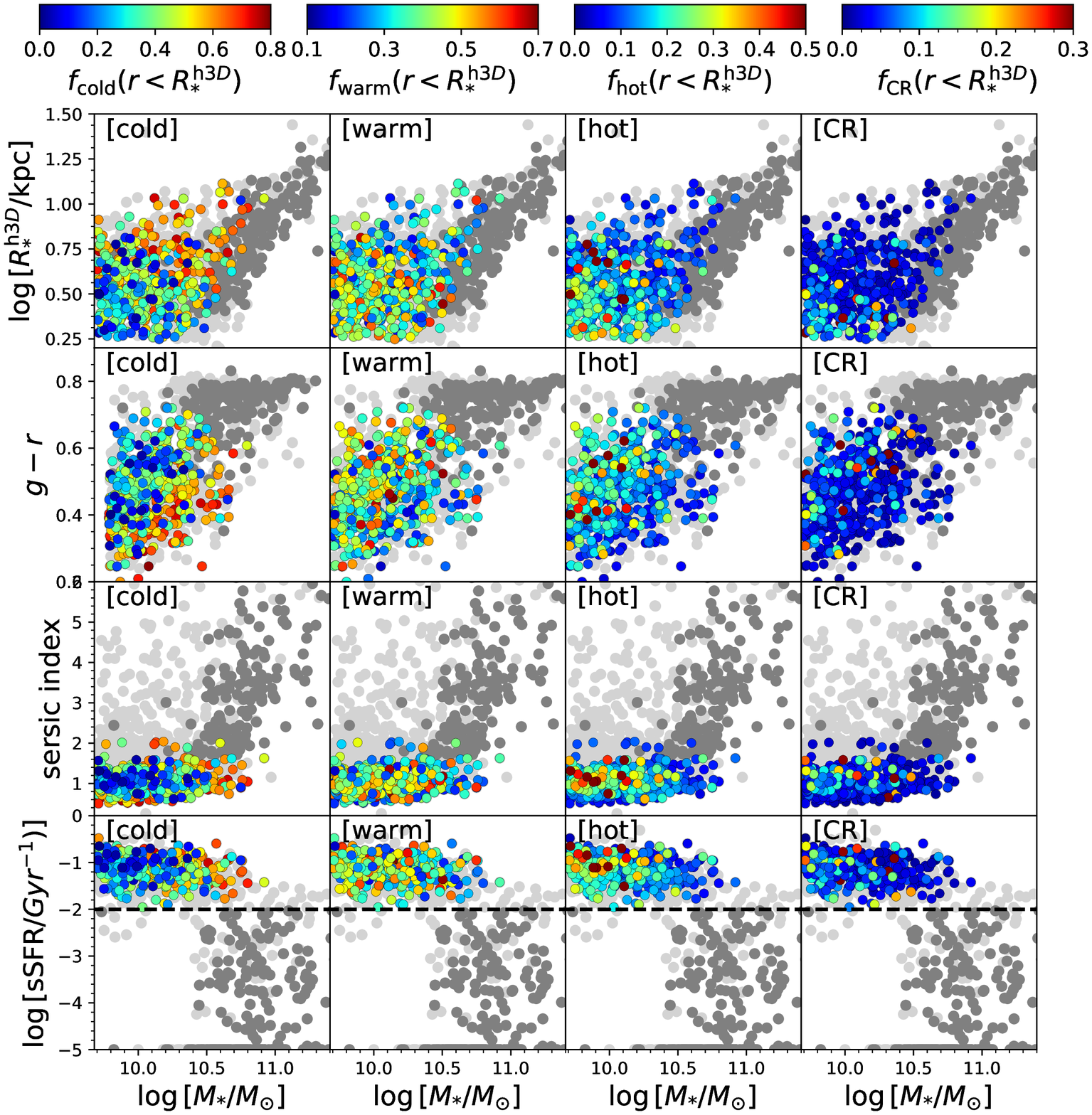}
\caption{From top to bottom we show galaxy size $R_{*}^{\rm h3 \it D}$, color $(g-r)_{\rm c}$, S{\'e}rsic index and sSFR as functions of stellar mass $\Mstar$, respectively for the TNG100 galaxies. Each light grey circle represents a central galaxy in the simulation sample. The dark grey circles represent explicitly classified early-type galaxies; while the colored circles represent late-type galaxies in the sample, whose orbital fractions of $f_{\rm cold}$, $f_{\rm warm}$, $f_{\rm hot}$ and $f_{\rm CR}$ are color-coded in columns from left to right, respectively. When plotting, late-type galaxies in {\it all} panels are ranked by their $f_{\rm hot}$ such that higher-$f_{\rm hot}$ galaxies appear on top of lower-$f_{\rm hot}$ ones. As a result, the red points (representing larger values) in the left-most column presenting the cold-orbit fractions appear as ``hidden'' behind the dark blue points (representing lower values). The dashed lines in the bottom panels correspond to $\log[{\rm sSFR}/{\rm Gyr}^{-1}]=-2$, which is used as one of the three criteria to define galaxy types. }
\label{fig:AllMassCodeFrac_LTG}
\end{figure*}

\begin{figure*}
\centering
\includegraphics[width=16cm]{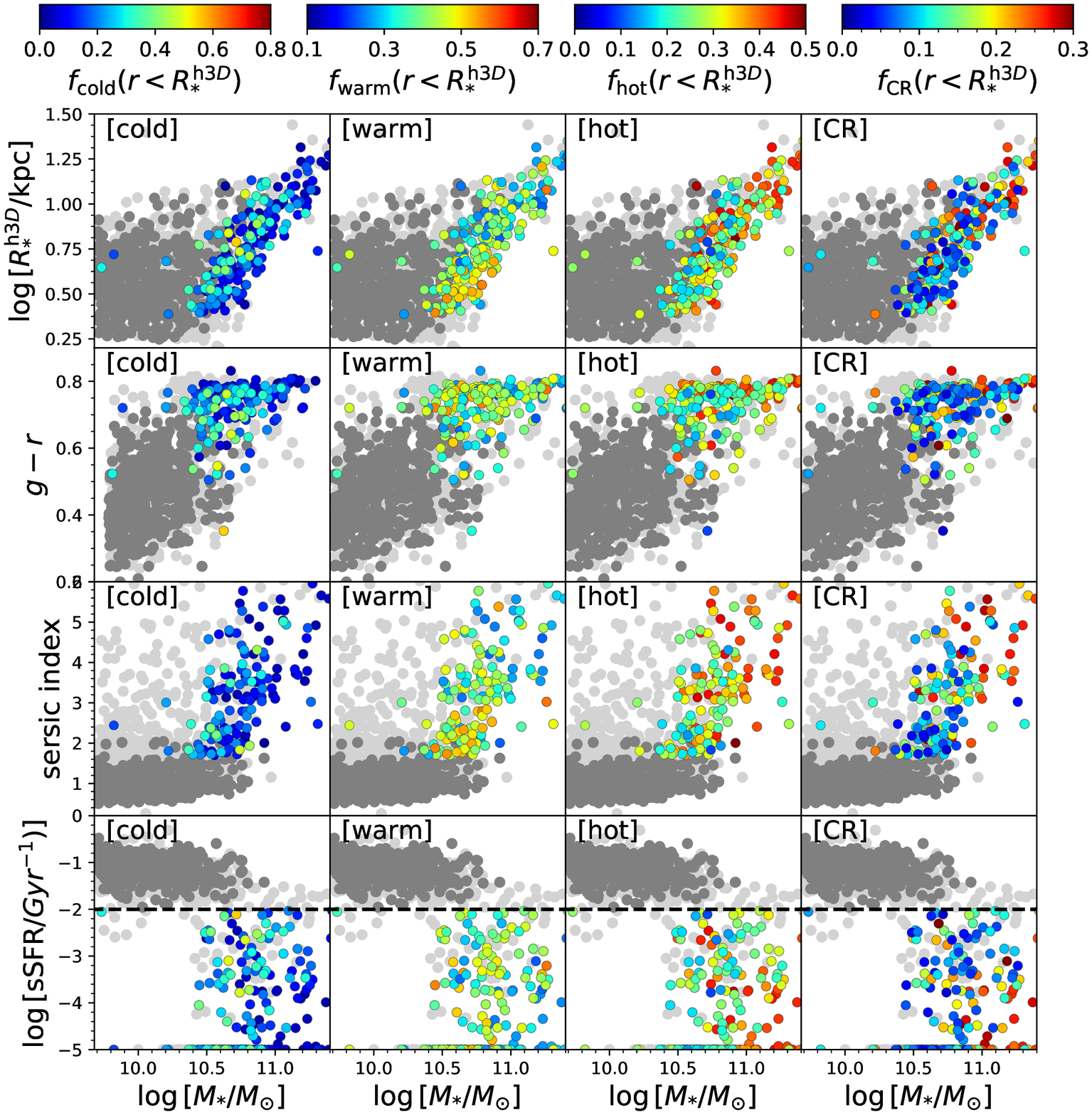}
\caption{Same as Fig.\,\ref{fig:AllMassCodeFrac_LTG}, except that the colored data points represent early-type galaxies, while dark grey circles are for their late-type counterparts. When plotting, early-type galaxies in {\it all} panels are ranked by their $f_{\rm cold}$ such that higher-$f_{\rm cold}$ galaxies appear on top of lower-$f_{\rm cold}$ ones. The dashed lines in the bottom panels correspond to $\log[{\rm sSFR}/{\rm Gyr}^{-1}]=-2$, which is used as one of the three criteria to define galaxy types.}
\label{fig:AllMassCodeFrac_ETG}
\end{figure*}

Using the method described in Sect.\,3.3, galaxies that are classified as typical late and early types occupy markedly different regions in the parameter space and have different stellar orbit compositions. This can be seen clearly from Fig.\,\ref{fig:AllMassCodeFrac_LTG} (for the late-type population) and Fig.\,\ref{fig:AllMassCodeFrac_ETG} (for the early-type population), where galaxy size $R_{*}^{\rm h3 \it D}$, color $(g-r)_{\rm c}$, S{\'e}rsic index and sSFR are plotted as functions of stellar mass $\Mstar$ from top to bottom, respectively. Each light grey circle represents a central galaxy in the simulation sample. In each figure for a given galaxy type, colored circles represent galaxies that are classified into the corresponding type; while dark grey circles represent galaxies that are classified into the opposite type for comparison. The orbital fractions $f_{\rm cold}$, $f_{\rm warm}$, $f_{\rm hot}$ and $f_{\rm CR}$ are color-coded in columns from left to right, respectively.

At stellar masses below $\Mstar \approx 3\times 10^{10}M_{\odot}$, the parameter space is mainly occupied by late-type galaxies, for which warm orbits dominate the kinematics, whereas $f_{\rm cold}$ is higher than $f_{\rm hot}$ on average (see Fig.\,\ref{fig:TNGvsCALIFA}). At an intermediate stellar-mass range between $3\times10^{10}M_{\odot}$ and $10^{11}M_{\odot}$, both early- and late-types galaxies exist: at a given stellar mass, galaxies with larger sizes, bluer colors, smaller S{\'e}rsic indices, and higher sSFR are identified as late-type galaxies; they also have larger $f_{\rm cold}$ and smaller $f_{\rm hot}$ than their early-type counterparts. Such correlations between the dynamical properties and the morphological, color and star-forming properties are expected by galaxy evolution theories. At stellar masses above $\Mstar \approx 10^{11}M_{\odot}$, the parameter space is largely occupied by early-type galaxies, whose kinematic compositions are dominated by hotter and counter-rotating orbits (which have perturbation origins); while cold orbits contribute the least. This is consistent with the finding that galaxy mergers play a dominant role in galaxy type transitions in most massive galaxies (\citealt{2017MNRAS.467.3083R, 2019arXiv190412860T}).

We point out, in particular, that above $\Mstar \approx 10^{10.5}M_{\odot}$, as galaxy mass increases, more and more galaxies transition from late types to early types, accompanied by reddened colors, increased S{\'e}rsic indices and diminished SFRs. Such a transition is also encoded by changes in the stellar orbit compositions such that more massive galaxies have lower $f_{\rm cold}$ and higher $f_{\rm hot}$ and $f_{\rm CR}$. Such changes in galaxies' kinematic compositions are an emergent result of the hierarchical structure assembly process under the framework of the cold dark matter cosmology and are also similar to those exhibited by the CALIFA-kinematic galaxies in Zhu18.

\section{Discussion and Conclusions}

In this work, we present a first study of the stellar orbit compositions of galaxies from the TNG100 simulation (\citealt{2018MNRAS.475..624N, 2018MNRAS.475..648P, 2018MNRAS.480.5113M, 2018MNRAS.477.1206N, 2018MNRAS.475..676S}). The goal of this study is to compare the stellar orbit contents between the simulated and observed galaxies, and to demonstrate the connections between fractions of different orbital components and other galaxy properties, such as mass, size, color, morphology and star formation rate.

To this end, we have selected a sample of $z=0$ TNG100 galaxies that have stellar masses of $\Mstar/M_{\odot}\in[10^{9.7},\,10^{11.4}]$. These selected simulated galaxies are comparable in redshift, mass, size and luminosity to the 260 CALIFA galaxies from \citet{2018NatAs...2..233Z}, for which an orbit-based Schwarzschild modelling technique (\citealt{1979ApJ...232..236S, 1993ApJ...409..563S}) was applied to the spatially resolved spectroscopic data. The stellar orbit compositions in terms of luminosity fractions of different orbital types were derived. This observational sample enables statistical analyses and theoretical comparisons over wide ranges of galaxy mass and morphology. In this work, we have resampled the stellar orbits from the best-fit Schwarzschild models at a given snapshot and calculated the {\it instantaneous} circularity distributions to classify four different orbital types of cold, warm, hot and counter-rotating (Sect.\,3.1). For the simulated galaxies, we have calculated the luminosity fractions of the four orbital types classified using the same criteria as for their observational counterparts (Sect.\,3.2).

We have found that the TNG100 simulation broadly reproduces the observed stellar orbit compositions and their stellar mass dependence. In particular, the mean distributions of the warm and the hot orbital fractions are well reproduced within model uncertainties of observed galaxies. Equally-well reproduced are the averaged fractions for the cold and counter-rotating orbits above and below $\Mstar \approx 6\times 10^{10}M_{\odot}$, respectively. In addition, the observed peak and trough features at $\Mstar \approx 1-2\times 10^{10}M_{\odot}$ in the mean distributions of $f_{\rm cold}-\Mstar$ and $f_{\rm hot}-\Mstar$, respectively, are also largely reproduced (see Fig.\,\ref{fig:TNGvsCALIFA} and Sect.\,4.1). Such a trend indicates a significant amount of ordered rotation starts to be replaced by ``random'' motion in higher and lower massive systems, consistent with the kinematic compositions of the CALIFA galaxies derived in Zhu18. We note that such a mass dependence was also revealed by the kinematic spheroid-to-total ratio and the stellar morphology concentration of the TNG100 galaxies, as studied by \citet{2019arXiv190412860T}.

Marginal disagreements are seen for the cold and the counter-rotating orbital component below and above $\Mstar \approx 6\times 10^{10}M_{\odot}$, respectively, where the average fractions are systematically higher than the observational data by $\la 10\%$ (absolute orbital fraction). The simulation also seems to overpredict and underpredict the scatter for the cold- and non-cold orbit fractions, respectively; in particular a good fraction of simulated galaxies at lower masses (mainly of late-types) are predicted to have much higher cold-orbit fractions in comparison to their observational counterparts.

Regarding the kinematically cold orbits, they are born as a consequence of in-situ star formation in a gaseous disk. With time, such orbits are gradually heated up due to secular evolution or galaxy mergers as they are the most susceptible to dynamical processes. We speculate that a higher predicted fraction of the cold orbits at lower masses may indicate an excess of cold gas disks unbalanced by insufficient heating at such mass scales. In the same direction, the larger (smaller) scatter in the cold-orbit (hotter-orbit) fraction may also indicate that either the duration of the involved heating mechanisms are not yet long enough, or the adopted heating models are not as diverse as in reality, and particularly so in late-type galaxies. For instance, heating effects from cosmic rays that can operate on large timescales are not included in the TNG simulation.

The simulation also overpredicts the mean fraction of counter-rotating orbits (by $\la 10\%$, absolute orbital fraction) in relatively high-mass galaxies, which are assembled earlier than their lower-mass counterparts. The formation of counter-rotating orbits, which are composed of older stellar populations, is thought to happen at a much earlier stage of galaxy evolution, during which both prograde and retrograde mergers contribute nearly equally to galaxy bulge growth. An excess of the counter-rotating fraction may indicate some prolonged epoch of such a process before the galaxy halo effectively accretes ambient cold gas, which cools and eventually forms a well-established gaseous and stellar disk at a later stage (see e.g., \citealt{2013MNRAS.430.2622D, 2019arXiv190205553P}). Such a delay may be caused by (1) a slower halo assembly process such that it has taken a bit longer than what is needed for the halo's potential well to grow deep enough; (2) some mechanisms that keep the ambient gas hot for a bit longer than what it should be before it can cool. The former process is less likely to be impacted by baryonic physics as it is mainly a consequence of dark matter halo assembly in the standard cosmological framework. The latter, however, could be affected by many factors in a simulation.

Galaxy kinematic composition reveals crucial and complementary information about galaxy evolution and in particular galaxy type transition (e.g., \citealt{2019arXiv190412860T, 2019arXiv190205553P}). To see this, we have further selected a late- and an early-type galaxy sample and compared differences in orbit compositions between the two galaxy types. As revealed by Figs.\,\ref{fig:AllMassCodeFrac_LTG} and \ref{fig:AllMassCodeFrac_ETG}, as galaxy mass increases, a higher fraction of galaxies are classified as early types; such a transition, accompanied by reddened colors, increased S{\'e}rsic indices and diminished SFRs, is also encoded in changes in the stellar orbit compositions such that the more massive galaxies are, the lower and higher the fractions of the kinematically cooler and hotter orbits become, respectively. Such changes in stellar kinematic structures with respect to galaxy mass are an emergent result of the hierarchical structure assembly process under the framework of the cold dark matter cosmology. 
%and are also exhibited by the CALIFA-kinematic galaxies in \citet{2018NatAs...2..233Z}.

It is interesting to notice that at stellar masses below $\Mstar \approx 3\times 10^{10}M_{\odot}$, where the majority of galaxies are identified as late types, a good fraction of them (most easily seen as the brighter-colored data points in the top row, third column in Fig.\,\ref{fig:AllMassCodeFrac_LTG}) show reversed relations between the cold- and the hot-orbit fractions, i.e., $f_{\rm cold} < f_{\rm hot}$, indicating the presence of a substantial amount of dynamical or perturbative heating. Such ``hot'' disks are also present in the observational sample (see Figure 2 of \citet{2018NatAs...2..233Z} for an example Sc-type galaxy with a significant hot-orbit component). The question is what has heated up these disks? Both a secular evolution and environmental perturbations from minor mergers could be the causes \citep{2016MNRAS.459..199G}. As can be seen from the figure, this unusual galaxy population is composed of galaxies with smaller sizes, as well as redder colors (indicating older stellar ages) at a fixed galaxy mass. It is worth noting that correlations between galaxies' kinematic composition and their color/age and size properties for the TNG100 galaxies were previously noticed: \citet{2019arXiv190412860T} found that early-formed TNG100 galaxies tend to have higher kinematic spheroid-to-total ratios; \citet{2018MNRAS.474.3976G} also found that quenched galaxies tend to be systematically smaller than their main sequence counterparts at fixed galaxy mass, similar to the trend noted from observations. It is likely that these ``hot'' late-type galaxies are on the way to be quenched through either dynamical heating or perturbative processes, which however, have not yet erased their disk morphologies (as seen from their small S{\'e}rsic indices).

It is also worth noting that at an intermediate stellar-mass range between $3\times10^{10}M_{\odot}$ and $10^{11}M_{\odot}$, a number of identified early-type galaxies (most easily seen as the brighter-colored data points in the top-left panel in Fig.\,\ref{fig:AllMassCodeFrac_ETG}) have fairly large $f_{\rm cold}$, comparable to what is seen for their lower-mass late-type counterparts, despite of their already established compact light morphologies (as seen from their large S{\'e}rsic indices). This indicates a coexistence of elliptical morphologies and kinematic disks with ordered rotation (still established at outskirts of galaxies, e.g., see Figure 2 of Zhu18 for an example of an early-type galaxy in the observational sample), which presumably happens during galaxy type transition in an ``inside-out growth'' fashion (e.g., \citealt{2009ApJ...697.1290B, 2010ApJ...709.1018V, 2010MNRAS.405.2253C, 2013AJ....146...77P}).

The findings above suggest that changes of a galaxy's morphology may not always be accompanied by changes in its kinematic composition; the mechanisms that drive galaxy type transition may be different at different mass scales (also see \citealt{2019arXiv190412860T}). We emphasize that such studies have become possible through combining information on stellar kinematic compositions and other fundamental properties for both simulated and observed galaxies. As we have seen here, this approach adds powerful new constraints on theoretical galaxy formation models.

\section*{ACKNOWLEDGEMENTS}

%DDX, RG and VS would like to thank the Klaus Tschira Foundation. 
DDX would like to acknowledges the TAP group at the Heidelberg Insititute for Theoretical Studies, the Center for Astrophysics at Tianjin Normal Universities (Xiaoyang Xia, Caina Hao), the stellar dynamic group at the Shanghai Observatories (Juntai Shen, Joao Antonio Amarante), as well as Peter Schneider, Dominique Sluse and Jingjing Shi for many insightful discussions. This work is partially supported by a joint grant between the DFG and NSFC (Grant No. 11761131004), the National Key Basic Research and Development Program of China (No. 2018YFA0404501), and grant 11761131004 of NSFC to SM. LZ acknowledges support from Shanghai Astronomical Observatory, Chinese Academy of Sciences under grant NO.Y895201009. GvdV acknowledges funding from the European Research Council (ERC) under the European Union's Horizon 2020 research and innovation programme under grant agreement No 724857 (Consolidator Grant ArcheoDyn). YW acknowledges fundings from NSFC (Grant No. 11773034, 11390372). FM is supported by the Program ``Rita Levi Montalcini'' of the Italian MIUR.

\bibliographystyle{mnras}
\bibliography{ms_xudd}
\label{lastpage}

\end{document}